\documentclass[aps,prl,amsmath,amssymb, twocolumn]{revtex4-2}

\usepackage{epsfig}
\usepackage{graphicx}
\usepackage{textcomp}
\usepackage[english]{babel}
\usepackage{xcolor}
\usepackage{soul}
\usepackage{bbding}
\usepackage{amssymb}
\begin{document}


\title{Water Drop on Thin Viscous Oil Layers: From Stick-Slip Spreading to Dewetting }

\author{Shubham Kumar$^\dagger$, Piyush Sahu$^\dagger$, Surjit Bharatsingh$^\dagger$, Gaurav Salwan}
\author{Dileep Mampallil}
\email[]{dileep.mampallil@iisertirupati.ac.in}

\affiliation{Indian Institute of Science Education \& Research Tirupati, Yerpedu P. O. PIN 517619, Tirupati, AP, INDIA}

\date{\today}

\begin{abstract}
The impact of water droplets on thin layers of immiscible viscous liquids, such as oil films, is commonly encountered across contexts ranging from kitchen activities to industrial processes. In this study, we experimentally investigate the short-term and long-term behavior of water drops spreading on silicone oil-coated surfaces. During the initial spreading, especially towards zero impact energies, the drop edge exhibits stick-slip dynamics, characterized by intermittent stops. The stick-slip behaviour diminishes with increasing spreading energy from impacts, where the drop spreads without noticeably displacing the oil layer.
In the long-term dynamics, regardless of whether the spreading is gradual or impact-driven, the drop eventually spreads onto the surface under the oil layer, governed by the dewetting dynamics of the oil. The delay for the second spreading is independent of the Weber number, indicating that the impact initially does not significantly deform the oil layer. Our findings provide new insights into the dynamics of water-oil interactions, with implications for both practical applications and fundamental research.
	
\end{abstract}
\maketitle
$^\dagger$ Equal contribution

\section{Introduction}

Drop impact and spreading are common phenomena observable in household and natural settings, human activities, and industrial processes. To list a few, processes, such as inkjet printing \cite{GDMartin2008}, drop-impact printing \cite{Prosenjit2020}, cooling systems \cite{Park2023}, spraying of pesticides and medicines \cite{MASSINON201765}, forensic analyses \cite{Attinger2013, Smith2018}, and natural events like rainfall \cite{SooNatCom2017} involve the dynamics of drop spreading and impact. Consequently, fundamentals of drop spreading \cite{Bonn2009} and impact \cite{Prosperetti1993, Yarin2006, Josserand2016, Liang2016, Visser2015} have been studied in various conditions. 

The spreading and impact dynamics of a liquid drop contacting a surface depend on factors such as the physical properties of the surface (solid, liquid, or viscoelastic), wettability, and the liquid's capillary and viscous properties \cite{Bonn2009, Visser2015, Yarin2006, Josserand2016}. The spreading dynamics are typically characterized by the evolution of the base radius ($R_b$), which often follows a power law of the form $R_b \propto t^\alpha$, where the exponent $\alpha$ varies based on the spreading conditions \cite{Eddi2013, Thampi, PRF104006}. For example, on solid surfaces, the inertial-capillary regime—associated with short-time dynamics—follows $\alpha = 1/2$ scaling \cite{Eddi2013, JiayuPoF}. In contrast, during the viscous-capillary regime, different exponents have been observed \cite{Eddi2013, Thampi}, with spreading dynamics resembling those of coalescing viscous drops \cite{Eddi2013}.

For drop spreading on liquid surfaces \cite{RAHMAN2018143, PRF104006}, a balance between viscous forces and surface tension predicts an exponent of $1/8$ \cite{joanny-hal56}. When the liquid surface is sufficiently thick, the spreading dynamics are governed by the interplay between interfacial tension gradients and viscous stresses, resulting in an exponent of $3/4$ \cite{Berg2009PoF, Matar2011PoF}. Various other exponents, ranging from $1/10$ to $3/4$, have been reported under different conditions. In cases involving very thin liquid layers, the spreading dynamics in the final viscous regime approach an exponent of $1/10$, which aligns closely with Tanner's law predictions \cite{Tanner_1979, Eddi2013}.

Although most studies on drop spreading focus on softly placed ones, drops often come into contact with surfaces through impact. Water drops impacting on oil layer is a common example, often encountered in kitchen activities or other industrial processes. In the case of drop impact on immiscible liquid layers, the interplay between inertial, capillary, and viscous forces governs the post-impact dynamics \cite{Fedorchenko2004}. The studies of water drop impact on oil layers \cite{Matar2018} and oil-infused surfaces \cite{Lee2014Langmuir, Kim2016Langmuir, Muschi2018} reveal that both the impact energy and the viscosity of the oil layer influence the behavior of the drop and the oil dynamics \cite{Matar2018, Lee2014Langmuir}. Lee et al. \cite{Lee2014Langmuir} found that oil viscosity does not significantly affect the maximum spreading radius, a key metric for characterizing drop impact under various conditions. However, at lower oil viscosity, the likelihood of bouncing increases \cite{Kim2016Langmuir}, whereas high-impact velocities result in crown formation \cite{Matar2018}. These findings suggest that the interaction between water drops and oil layers, or more generally, the impact of drops on liquid layers, involves complex dynamics that require further investigation.

In this experimental study, we investigate the spreading dynamics of water drops on silicone oil layers coated on various substrates, with and without impact, revealing several previously unreported dynamics. We observe that the drop spreads in a stick-slip manner, with the effect becoming more pronounced as the oil viscosity increases. Upon impact, a slight increase in the spreading velocity diminishes the stick-slip behavior. In both pure spreading and impact scenarios, the drop eventually spreads over the underlying substrate after a delay, which is governed by the dewetting dynamics of the oil layer beneath the drop. Notably, we find that this delay time is independent of the impact velocity, contrary to the expected rupturing the oil layer immediately upon impact. Through systematic experiments, we explore the different spreading dynamics exhibited by the impacted drops and compare our results with established scaling laws, shedding new light on the interplay between impact energy, oil viscosity, and interfacial behavior.

\section{Experiments}

We spin-coated silicone oil on glass slides sequentially cleaned with isopropyl alcohol and water, followed by air drying. The spin coating of silicone oil was performed at different rpm ranging from 500 to 5000. Different rpm resulted in different thicknesses of the oil, depending upon the oil viscosity (kinematic, $\nu_o$), such as 370 and 10000 cSt (Sigma Aldrich). The oil thickness was obtained by comparing the weight of the glass slide in a microbalance (Quintix224-10IN) before and after the uniform coating with the oil. The thickness was obtained as $d_o = \delta m /\rho_o A$, where $\delta m$ is the oil mass, $\rho_o$ is the density of oil, and $A$ is the area of the glass slide. An average thickness was obtained from five different samples. The relationship between the spin rpm and the oil layer thickness for different oils is shown in Supplementary Fig.S1. We compared the measured oil thicknesses to the ones obtained using the Meyerhofer model \cite{Meyerhofer}, which gives the spin-coated thickness as $(3\nu_o/4\omega^2 T)^{1/2} $, where,  $\omega$ is the angular velocity of spinning and $T$ is the duration of the spinning. The calculated values were higher than the measured ones up to two times, especially at lower rpm. We used the experimental values of the thickness for further analyses.  

We coated oil on different substrates such as glass, indium tin oxide (ITO), and polydimethylsiloxane (PDMS). On these bare surfaces (without oil coat), water drops have equilibrium contact angles of 22 $\pm$ 2$^\circ$, 60 $\pm$ 2$^\circ$. and 100 $\pm$ 2$^\circ$, respectively.

Water drops of volume 6.5 $\mu$l (with 5\% deviation) having radius $R$ = 1.1 mm were dropped from a Teflon tubing connected to a syringe pump. The impact height varied from 0 to 0.95 m. The corresponding impact velocity varied between 0 and 4.3 m/s. Upon impact, the dynamics of the drop were imaged using a high-speed camera (Phantom MICRO LAB110) at frame rates up to 10000. The images were analyzed using ImageJ software and homemade Matlab code to obtain the contact angle and base radius as a function of time.

\begin{center} 
	\begin{figure*}[ht]
		\centering
		\includegraphics[scale=1.35]{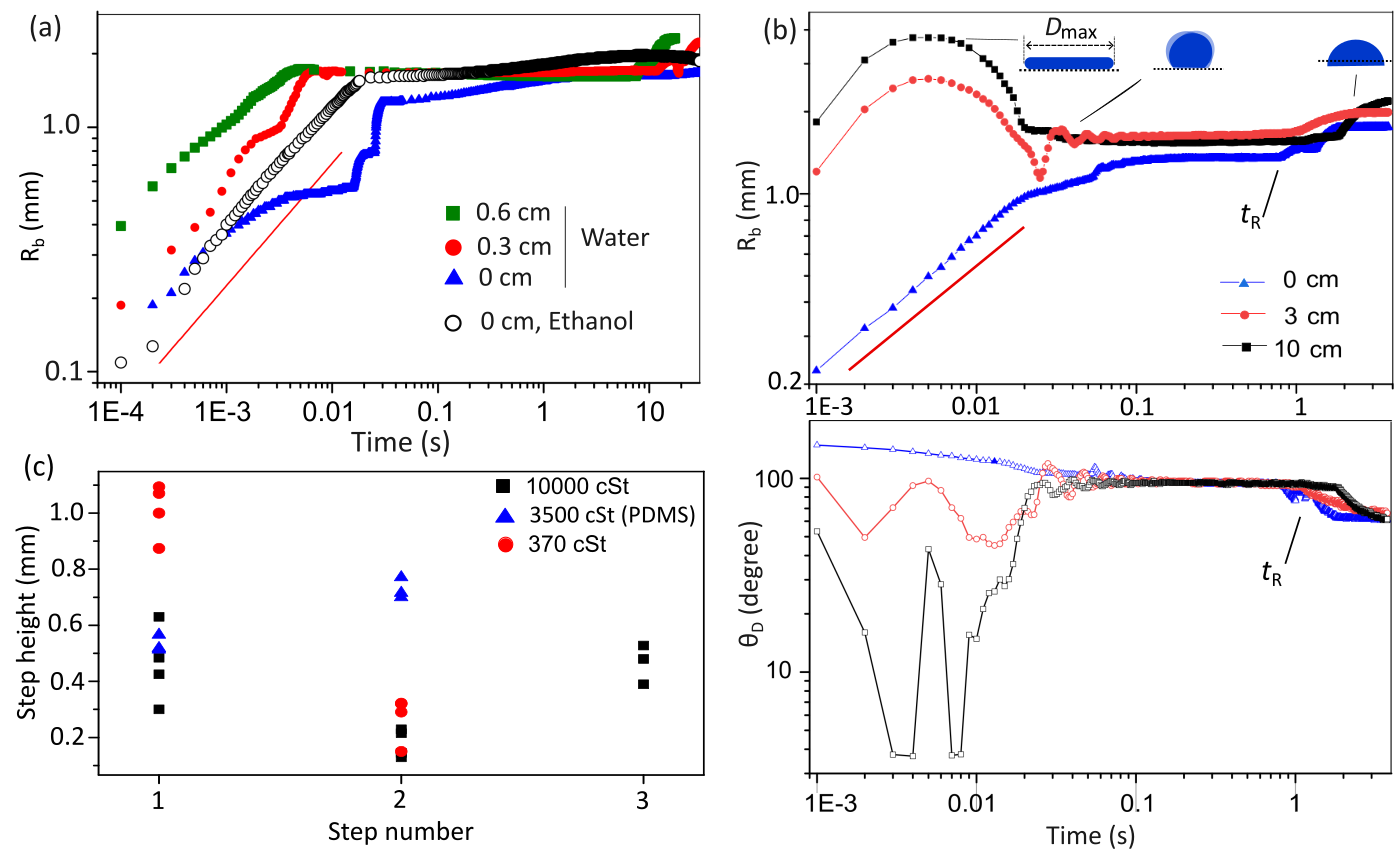}
		\caption{ (a) The base radius ($R_b$) of water drops on a 10k cSt silicone oil surface. When gently placed, the spreading follows a step-like progression, transitioning to a smooth dynamic upon impact. Ethanol drops lack step-like behavior. The solid line has a slope of 1/2. (b) $R_b$ of drops placed gently and impacted from two different heights (3 cm and 10 cm) on a 370 cSt silicone oil layer, 7 $\mu$m thick. In the non-impact case, a step occurs near the end of spreading. The solid line shows a slope of 1/2. Under impact, after reaching maximum spread, the drop undergoes weakly damped oscillations before stabilizing with a constant $R_b$ and dynamic contact angle ($\theta_D$). During this stable phase, the drop dewets the oil layer. At the end of dewetting, the drop spreads on the underlying surface at $t = t_R$, marked by a reduction in the contact angle as the contact line reaches the substrate. (c) Step heights in the $R_b(t)$ curve for gently placed water drops on silicone oil and uncured PDMS layers. The first step is relatively large, while the final step completes spreading on the oil.  }\label{Fig1}
	\end{figure*}
\end{center}

\begin{center} 
	\begin{figure}[ht]
		\centering
		\includegraphics[scale=0.95]{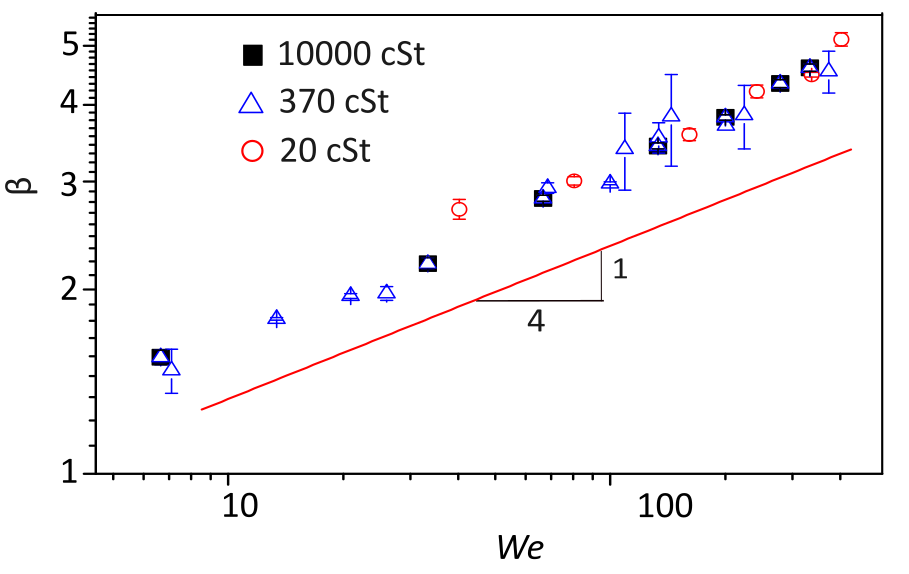}
		\caption{ The maximum spreading parameter $\beta = D_{max}/D$ as a function of Weber number. The error bars are the standard deviation from at least five measurements. The inset illustrates a drop spreading on the oil layer with various parameters marked.  }\label{Fig2}
	\end{figure}
\end{center}

\begin{center}
	\begin{figure*}[ht]
		\centering
		\includegraphics[scale=0.99]{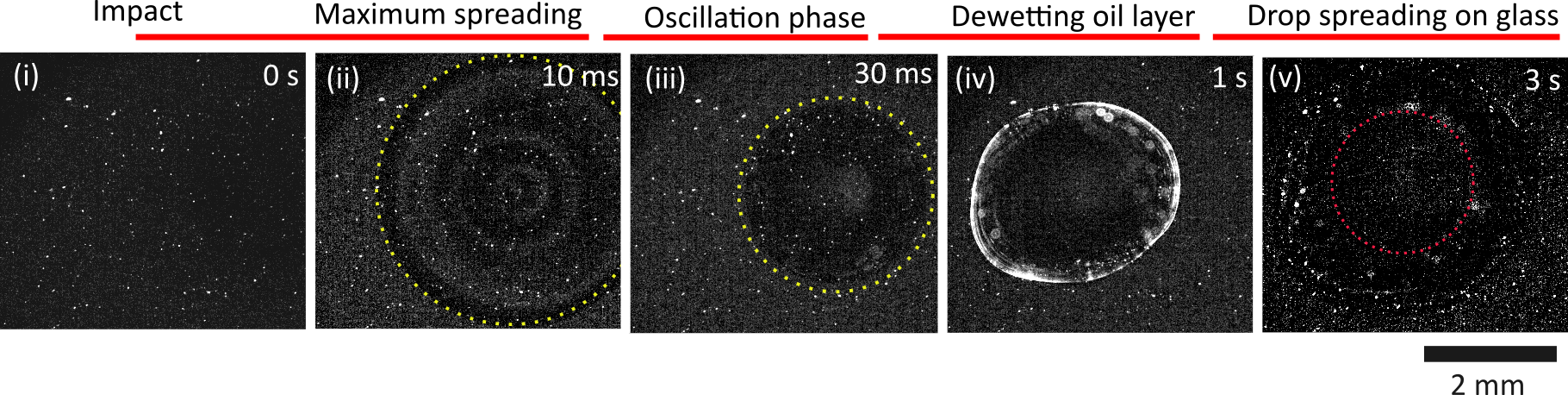}
		\caption{ Images showing the impact dynamics of a drop on an oil layer (viscosity: 370 cSt) containing suspended fluorescent particles (diameter 2 $\mu$m). (i)-(ii): From impact to maximum spreading, no particles are displaced from their original position, implying the integrity of the oil layer during the drop spreading. (ii)-(iii): Only a few particles are dragged radially inward during the retraction phase. The yellow dashed circle represents the edge of the water drop. (iv)-(v): After the dewetting process under the drop, the drop spreads on the underlying glass surface. The red dashed circle in (v) represents the oil droplet trapped at the center.  }\label{Fig3}
	\end{figure*}
\end{center}

\section{Results and discussion}

Upon contact with the oil layer, the water drop rapidly spreads on the layer.  The overall spreading process consists of two distinct phases: (i) an initial rapid spreading phase, followed by (ii) a stable region where the oil under the drop dewets, and (iii) a delayed spreading phase that onset at a time marked $t_R$. In the following sections, we will analyze each phase in detail.

\subsection{Initial dynamics: stick-slip spreading}

First, we gently place a water drop on the oil layer of viscosity 370 cSt and 10k cSt, i.e., without impact. The drop spreads rapidly with an initial scaling $t^{1/2}$, independent of the oil’s viscosity (Fig.~\ref{Fig1}(a) \& (b)). This scaling is consistent with the spreading dynamics of low-viscosity drops on solid surfaces \cite{Eddi2013}. It indicates that during the initial rapid spreading, any viscous dissipation at the drop’s contact line is negligible.

As the spreading progresses, the $R_b(t)$ curve exhibits step-like variations, indicating a stick-slip behavior. This effect is more prominent when the drop spreads on higher-viscosity oil (10k cSt) compared to that on 370 cSt.

We propose that the step-like behavior arises from the advancing contact line being obstructed by the oil layer. The contact line rapidly moves outward by a force $f_s =  \gamma(\cos\theta_E - \cos\theta_D(t))$, where $\theta_E$ is the equilibrium contact angle and $\theta_D(t)$ is the dynamic contact angle. The spreading contact line likely pushes a thin layer of oil, forming a ridge along its edge. As the ridge grows, the viscous force $f_v$ also grows. Consequently, the contact line stops the movement beyond $f_s = f_v$. Similar slowing down of the contact line occurs for drops spreading on soft viscoelastic surfaces \cite{Tamim_Bostwick_2023}. In our system, the contact line stops much earlier than the contact angle becoming the equilibrium value on the oil. As a result, the dynamic contact angle increases. Thus, the drop edge flows over the ridge and resumes spreading. This cycle of obstruction and release produces the characteristic stick-slip pattern observed in the $R_b(t)$ curve.

Low-viscosity oil (370 cSt) presents less resistance to the advancing contact line, resulting in only one step with a less distinct plateau before the spreading finally stops at $\theta_D = \theta_E$. Also, the step appears towards the end of the spreading phase. In contrast, high-viscosity oil causes two pronounced steps before the drop reaches the condition $\theta_D = \theta_E$. This difference arises because the low-viscosity oil is easily displaced by the spreading drop, allowing for relatively smoother outward movement. However, with high-viscosity oil, the drop faces greater resistance, causing repeated stops and jumps as it struggles to push through the thicker oil layer.

Consistently, ethanol drop spreading on the oil layer exhibited no stick-slip dynamics. Since ethanol is less dense (0.789 g/cm$^3$) compared to the oil (0.968 g/cm$^3$), no edge penetration occurred into the oil. As a result, the spreading of ethanol drop is smooth and follows the scaling $R_b \propto t^{1/2}$ (Fig.~\ref{Fig1}(a)).

The stick-slip spreading observed in our experiments resembles the behavior of drops spreading on thin powder (soot) layers coated on solid surfaces \cite{PRE.98.043107}. As the contact line advances on soot, it penetrates the layer, pushing particles into concentric circular ridges with progressively smaller ridge-to-ridge separations \cite{PRE.98.043107}. In our case, the first ridge (step on $R_b(t)$) forms farther out because the rapid initial spreading velocity allows the drop to glide over the oil surface without immediately pushing much oil outward—similar to the quick spreading observed on thin powder layers \cite{PRE.98.043107}. The second step (for 10k cSt oil) is smaller because it occurs after the drop has slowed down, following partial spreading. The third step for 10k CSt and the second step for 370 cSt oils do not correspond to ridge formation but mark the completion of the drop’s spreading on the oil surface. Since the dynamic contact angle is already small at this stage, we assume that viscous dissipation during the final step is minimal. We have quantified the step height in Fig.~\ref{Fig1}(c). It shows that with decreasing oil layer viscosity, the height of the first step increases, indicating that it appears at a later time of the spreading.


When impact is involved, the stick-slip behavior diminishes, and the steps in the spreading curve disappear (Fig.~\ref{Fig1}(a)). We assume that a faster outward spreading reduces the obstruction from the accumulated oil layer, allowing relatively smoother movement of the contact line.

The smoother spreading without obstructions is also reflected in the maximum spreading diameter of the drop upon the impact. The maximum spreading diameter is an important parameter as it has implications in several applications involving drop impact \cite{clanet2004, Pasandideh650, Eggers3432498, Ukiwe, Josserand2016, WORNER2023104528, PRA044018}. The maximum spreading is characterized by a factor, $\beta$, which is the ratio of the diameter upon maximum spreading ($D_{max}$) to that at the initial spherical drop ($D = 2R$). The value of $\beta$ results from the balance of kinetic energy, surface energy, and viscous dissipation. Considering these aspects, numerous relationships were derived connecting $\beta$ to dimensionless numbers such as Weber number, ${We} = \rho_w D u_i^2/\sigma_w$ and Reynolds number, ${Re} =  D u_i/\nu_w$, where, $\rho_w$ is the density of the impacting liquid (water),  $u_i$ is its impact velocity, $\sigma_w$ is the surface tension, and $\nu_w$ is the kinematic viscosity of the impacting drop. In the viscous regime where spreading occurs by a balance between the kinetic energy and the viscous dissipation,  $\beta \sim {Re}^{1/5}$ was reported \cite{Eggers3432498, Josserand2016}. When viscous effects are small, a comparison of the initial kinetic energy with the capillary energy has predicted  $\beta \sim {We}^{1/2}$ at asymptotic regimes of large Weber numbers. Clanet et al. derived another scaling law, $\beta \sim {We}^{1/4}$, considering the volume conservation between the initial spherical and the final pancake shapes \cite{clanet2004}. They considered that the final thickness of the pancake drop is proportional to the capillary length $a_c = \sqrt{\sigma_w/(\rho_w g')}$, where $g' \sim u_i/D$ is an increased acceleration experienced by the drop during the spreading. 

In our experiments, we observe the scaling  $\beta \sim {We}^{1/4}$, indicating smooth spreading without obstructions, unlike in the case of spreading without impact (Fig.~\ref{Fig2}).  Although the scaling ${We}^{1/4}$ was first demonstrated for low-viscosity drop impact on strongly water-repellent surfaces \cite{clanet2004}, impact on various surfaces \cite{Ukiwe, Josserand2016} including oil-infused surfaces \cite{Lee2014Langmuir} have shown this scaling.  Various other works reported different scaling laws as ${We}^{1/2}$, ${We}^{1/4}$, ${Re}^{1/4}$, ${Re}^{1/5}$ or complex relationships suggesting a universal scaling \cite{PRA044018, Lee-bonn2016, Josserand2016}. Since correlating the scaling nature with the experimental parameters is difficult, recently, machine learning has been employed to explore the maximal spreading dynamics \cite{Tsai2023}.

We observed the same scaling regardless of the oil layer viscosity. This behavior can be expected as the spreading is inertial and peripheral on the oil layer. It also explains the absence of step-like spreading during impact, as the fast-spreading drop does not disturb the oil layer. To confirm the peripheral spreading of water drop on the oil layer, we dispersed red fluorescent polystyrene particles of diameter 2 $\mu$m in the oil ($\nu_o$ = 370 cSt). We observed the impact from below the glass slide (Fig.\ref{Fig3}). Upon impact, the drop spreads without creating any noticeable movement of the particles indicating that the spreading is mostly peripheral. Also, the side view of the drop demonstrated an advancing angle of more than 90$^\circ$. After the maximum spread, during the retraction, the drop drags the oil radially inward manifested by the corresponding movement of the fluorescent particles. It demonstrates that the drop dissipates some of its energy in the oil during the retraction phase. Followed by retraction, the drop undergoes damped oscillations where it loses most of its kinetic energy \cite{Bandyopadhyay}.

In short, during the initial spread without impact, the water drop disturbs the oil layer peripherally, leading to a step-like motion of the contact line. This effect is more noticeable on high-viscosity oil, where the contact line faces greater resistance in displacing the oil. However, with impact, the drop’s higher kinetic energy overcomes this resistance, ensuring a smooth spreading. Other factors may also influence the dynamics of impacted drop spreading. For example, air film entrapment between the impacting drop and the oil layer \cite{RuiterPRL2012, BouwhuisPRL, maleprade2013} could lead to differences in spreading behavior on low- and high-viscosity oil surfaces due to their varying deformability. Additionally, air entrapment may leave the oil at the center of the impact point relatively undisturbed, as observed in the case of drop impact on granular media, where it forms distinct patterns \cite{Sharma2020}.

\begin{center}
	\begin{figure*}[ht]
		\centering
		\includegraphics[scale=0.99]{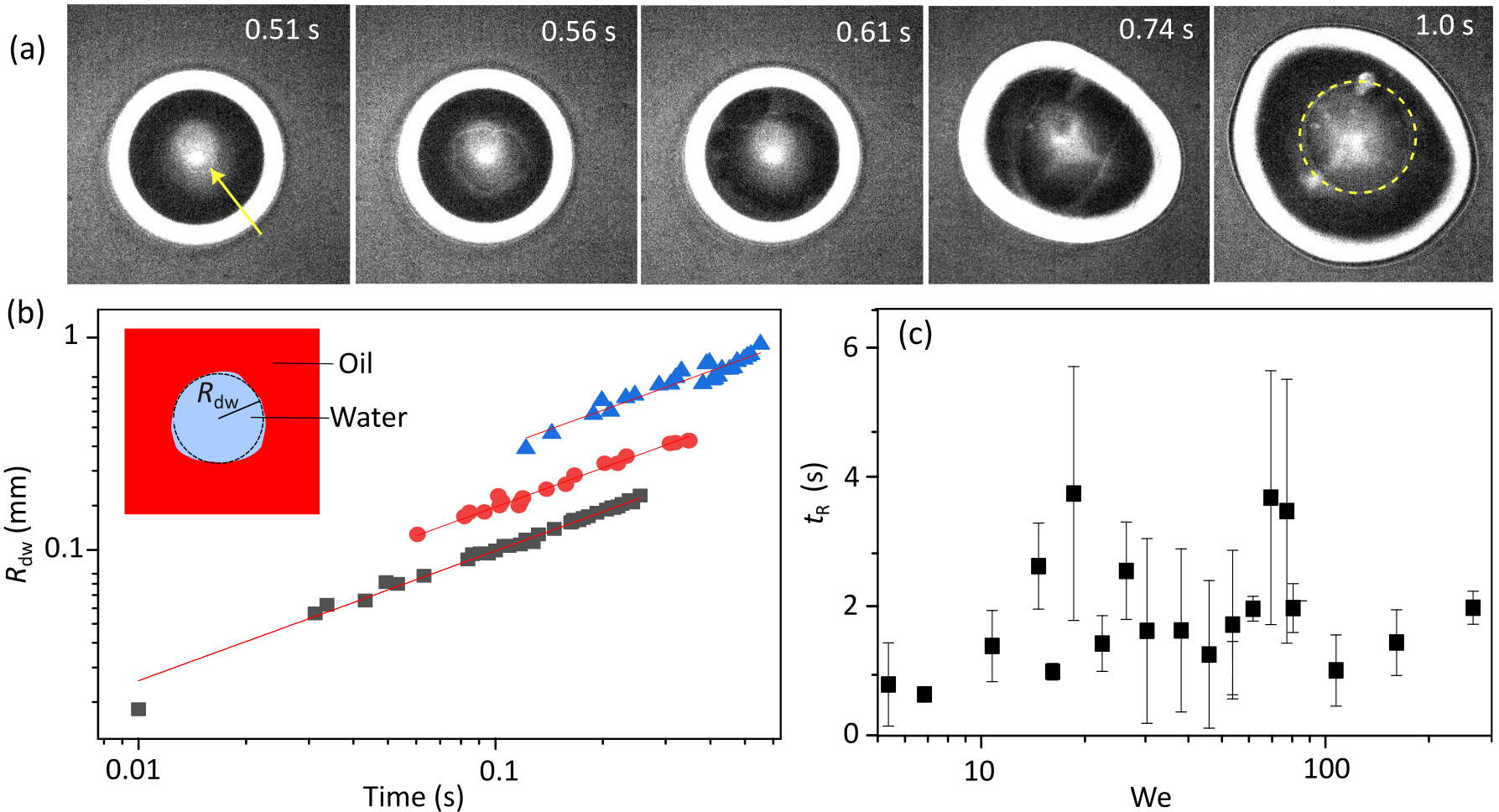}
		\caption{ 
			(a) About 0.5 s after impact, water ruptures the oil layer (marked by the arrow) and initiates the dewetting of the oil layer. Finally, under the spreading water drop, an oil droplet is formed (see inside the dashed circle on the panel of 1.0 s and Supplementary video 1). The bright circular periphery of the drop is due to light reflection. (b) The radius of the dewetting hole $R_{dw}$ over time for three measurements. The solid lines have a slope of 2/3. (c) The delay to begin the second spreading, $t_R$ is independent of the Weber number. It was measured from the side view of the drops. Error bars represent the standard deviation from at least three measurements. Random nucleation of the dewetting hole causes significant variation in $t_R$ when the hole comes into contact with the drop’s edge, triggering spreading.   } \label{Fig4}
	\end{figure*}
\end{center}

\begin{center}
	\begin{figure}[h]
		\centering
		\includegraphics[scale=0.99]{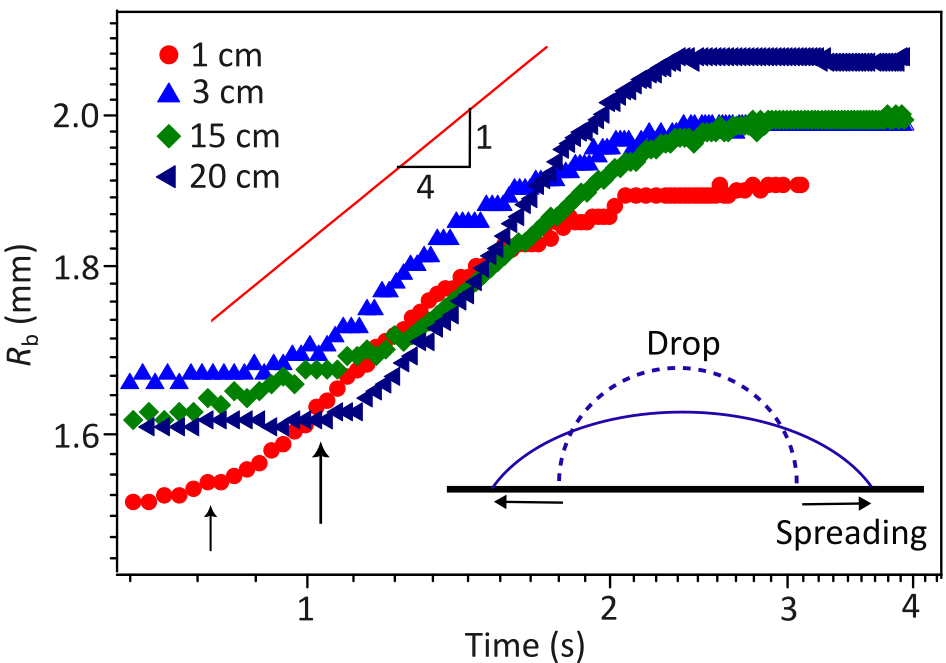}
		\caption{ 
			After impact, the drop spreads on the solid surface beneath the oil layer after a time $t_R$ (indicated by arrows). The slow increase in the base radius $R_b$ reflects the characteristics of viscous drop spreading, with varying slopes. For reference, a solid line with a slope of 1/4 is shown, indicating slower spreading compared to that on the oil layer, which had an exponent of 1/2. The impact heights are labeled in the panel. The inset schematic illustrates the initial (dashed line) and final (solid line) shapes of the spreading drop.  } \label{Fig5}
	\end{figure}
\end{center}

\subsection{Delayed dynamics: dewetting the oil layer}

After the gently placed drop completes its initial spreading or the impacted one completes its damped oscillation phase, it maintains a constant contact angle of approximately 100$^\circ$ in a quasi-equilibrium state. This state exists for a duration of $t_R$, after which the drop undergoes a second spreading characterized by an increasing $R_b$ and a decreasing contact angle to approximately 35$^\circ$ (Fig. 1b). We observed similar spreading of water drop also with PDMS and ITO substrates coated with silicone oil. When an oil layer (7 $\mu$m thick) was applied to PDMS and ITO surfaces, after a duration $t_R$, the final equilibrium contact angles were measured to be 100 $\pm$ 2$^\circ$ and 65 $\pm$ 2$^\circ$, respectively. These values closely align with the contact angles observed on the corresponding bare surfaces, namely 100 $\pm$ 2$^\circ$ for PDMS and 60 $\pm$ 2$^\circ$ for ITO. These findings strongly suggest that the change in contact angle occurs as the drop's contact line penetrates the oil layer and interacts with the underlying substrate. This is expected, as the oil density is slightly lower than that of water. 

On glass, the final equilibrium contact angle did not match the contact angle of water on the bare substrate. For example, while the drop displayed an equilibrium contact angle of 22$^\circ$ on bare glass, it exhibited a slightly larger angle (approximately 35$^\circ$) after spreading with an oil layer. This difference might be attributed to the potential presence of a thin oil film trapped beneath the spreading drop \cite{PRLStaicu, HONG2014292}. 

We further examined the water penetration into the oil layer. The bottom view of the drop reveals that, before spreading over the underlying substrate, the water ruptures the oil layer, initiating dewetting (Fig.~\ref{Fig4}(a) and Supplementary Video 1). During dewetting, a hole nucleates within the oil layer beneath the water drop. The dynamics of dewetting depend on several factors, including the substrate’s wettability, film thickness, viscosity, and surface tension of the liquids involved \cite{PRL088301, Baumchen2010, pnas.1820487116}. Studies show that dewetting on thin ($<$100 nm) PDMS layers occurs homogeneously through the nucleation of multiple holes, followed by their coalescence. In contrast, on thicker layers, dewetting occurs heterogeneously, with a single hole forming randomly beneath the drop \cite{BHATT2022530}. In our observations, we found that a single hole appears at a random location under the drop and grows over time. 

The radius of the dewetting hole is reported to follow a scaling law, $R_{dw} \propto t$, particularly for thicker films and non-slipping substrates \cite{GennesTextBook}. We found that the growth of the hole radius follows the scaling law $R_{dw} \propto t^{2/3}$, which is characteristic of a thin sliding oil layer \cite{RedonPRL1991, GennesTextBook}. As shown in Fig.\ref{Fig4}(b), the observed scaling has an exponent of 0.66 ± 0.03. This differs from the findings of Bhatt et al., who reported a linear scaling for PDMS film thicknesses ranging from 28 nm to 8000 nm on silicon wafers \cite{BHATT2022530}. In our case, the dewetting may leave a trail of precursor oil film, effectively forming a slipping surface. As the dewetting hole grows, water fills the hole above the precursor film. Upon completion of the dewetting, the oil film forms a small oil drop under the water drop (Fig.\ref{Fig3} Fig.\ref{Fig4}(a), and Supplementary Video 1).

In our experiments, the rupture and dewetting process begins only a few hundred milliseconds after impact. This brief delay can be attributed to the entrapped air layer between the water drop and the oil, similar to previous observations where air entrapment initially delays the engulfment of water drops by oil \cite{CuttleJFM2021}. Dewetting proceeds beneath the drop while it maintains a quasi-equilibrium contact angle, as observed from the side. Once dewetting is complete at time $t_R$, the drop begins to spread over the underlying substrate. The duration $t_R$ for dewetting to reach the drop edge (and for that region to start spreading) depends on the location where the initial dewetting hole nucleates. This duration is approximately one second on the 370 cSt oil layer and ranges from 10 to 50 seconds on the 10k cSt layer.

We further note that $t_R$ is independent of the impact parameters. Specifically, the value of $t_R$ remains unchanged with ${We}$, as demonstrated in Fig.~\ref{Fig4}(c). Dewetting begins only after the impact and the settling of oscillations. Therefore, it is expected that the dewetting process is not influenced by the history of the drop impact unless the impact disrupts the oil layer. The independence of $t_R$ from ${We}$ also supports our observation that the drop’s spreading and retraction are primarily peripheral and do not significantly displace the oil.

Following dewetting, the drop spreads across the substrate surface, exhibiting an increase in the base radius (see region after approximately 1 s in Fig.\ref{Fig1}(b)). This region is highlighted in Fig.\ref{Fig5}. The spreading behavior resembles that of a viscous drop, lacking a pure power-law pattern as noted in \cite{Eddi2013}. Overall, this spreading is slower than that of low-viscosity drops, which typically follow a power law with an exponent of $1/2$ \cite{Eddi2013}. We assume that despite our drop being a low-viscosity liquid (water), the resistance posed by the viscous oil affects the movement of the contact line. Moreover, the spreading regime does not align with early times, as the drop had already partially spread on the oil layer. Consequently, we anticipate a smaller exponent, indicative of the late-time dynamics of viscous droplets.

\section*{Discussion}

Our observations indicate that the drop spreading without impact involves some stick-slip-like spreading, possibly due to the formation of an oil ridge around the moving contact line. Notably, the stick-slip spreading is absent when the drop is less denser than the oil. Upon impact,  the spreading becomes smoother.  Moreover, the drop initially does not deform the oil layer contrast with the findings from other studies on water drop impact on oil \cite{Matar2018, Lee2014Langmuir, Kim2016Langmuir}. For instance, Che et al. \cite{Matar2018} examined water drop impact on silicone oil layers with viscosity ranging from 5 to 50 cSt. They noted that the impacted drop displaces the oil layer, but this effect diminishes with increasing oil viscosity, consistent with our findings. Our measurements, conducted at much higher viscosity (370 and 10k cSt), reveal that the impacted drop does not disturb the oil layer during its spreading phase.

In experiments with very low viscosity oil, such as our trials with 20 cSt silicone oil, the retracted water drop retained considerable kinetic energy, resulting in the formation of a vertically elongated jet shape (see Supplementary Fig. S2). At large impact energies, this jet subsequently broke up to form a small drop at the tip. This observation aligns with previous findings indicating that the bouncing of the impacted drop occurs when the oil viscosity is low \cite{Kim2016Langmuir}. However, these effects did not exist with a relatively higher viscous oil layer as the drop dissipated most of its kinetic energy, especially during retraction.

\section*{Summary}

In summary, motivated by the simple and familiar scenario of water drops falling on oil layers, we investigated the spreading dynamics of water drops on thin silicone oil layers coated on solid surfaces. In the absence of impact, drop spreading was not smooth but exhibited stick-slip dynamics. However, these effects were minimal during impact, as the contact line advanced rapidly on the oil layer. Upon impact, the drop did not significantly deform the oil layer, with smooth spreading occurring primarily on the oil surface. In both impact and non-impact spreading, the water drop dewets the oil layer beneath it, with the growth of the dewetted region following a $R_{dw} \propto t^{2/3}$ scaling. After dewetting is complete, the drop spreads on the underlying substrate similarly to a viscous drop spreading on a solid surface. Eventually, the precursor oil film retracts, forming a small oil droplet beneath the water drop as it spreads on the solid substrate. In short, we reveal previously unreported spreading dynamics of water drop spreading on oil layers. Our findings may inspire further experimental and theoretical research in related fields.

\section{Acknowledgment}
DM acknowledges IISER Tirupati intramural funds and Science and Engineering Research Board (India) grant CRG/2020/003117. PS acknowledges the financial support from CSIR, India.

\section{Competing financial interests}
The authors declare no competing financial interests.

\section{References}

\bibliography{ref_man_1}


\end{document}